\documentclass[aps,prb,twocolumn,letterpaper,superscriptaddress]{revtex4-2}

\usepackage{amsmath}
\usepackage{amssymb}
\usepackage{amsfonts}
\usepackage{bm}
\usepackage{mathrsfs}
\usepackage{tensor}
\usepackage{dsfont}
\usepackage{esint}
\usepackage{comment}
\usepackage{graphicx}
\usepackage{dcolumn}

\usepackage[usenames,dvipsnames]{xcolor}
\usepackage[hyperindex,pdftex, breaklinks,colorlinks = true,linkcolor = blue,urlcolor=blue,citecolor=blue]{hyperref}

\usepackage{physics}
\usepackage[section]{placeins}

\begin{document}
	
	\title{A new perspective on the anomalous Hall effect}

	\author{Jason G. Kattan}
	\email{jason.kattan@mcgill.ca}
    \affiliation{Department of Physics, McGill University, Montréal, Québec H3A 2T8, Canada}
	\affiliation{Department of Physics, University of Toronto, Toronto, Ontario M5S 1A7, Canada}

    \author{Matthew Albert}
    \email{matthew.albert@mail.utoronto.ca}
    \affiliation{Department of Physics, University of Toronto,
    Toronto, Ontario M5S 1A7, Canada}
 
	\author{J. E. Sipe}
	\email{john.sipe@utoronto.ca}
	\affiliation{Department of Physics, University of Toronto, Toronto, Ontario M5S 1A7, Canada}
	
	\date{\today}
	
	\begin{abstract}
             We revisit the anomalous Hall effect in magnetic conductors, and its generalization to finite frequencies, using a formalism based on microscopic notions of polarization, magnetization, and free charges and currents. The electronic degrees of freedom are treated within second-quantized field theory, where the Hamiltonian features a static and cell-periodic magnetic field that encodes the magnetic order in the crystal and breaks time-reversal symmetry. We study the dynamics of bound and free charge carriers at the microscopic level as they respond to a spatially uniform electric field at finite frequency. The conductivity tensor describing the long-wavelength response is a sum of three terms, including a Kubo term associated with the polarization response, along with the metallic Drude term and the anomalous Hall conductivity that are associated with the longitudinal and transverse parts of the free current response, respectively. We also present numerical calculations of these contributions for the ferromagnetic body-centered cubic phase of iron.
        \end{abstract}
	
	\maketitle


\section{Introduction}\label{Sec:Introduction}

The classical Hall effect has occupied a central place in the electrodynamics of materials ever since Edwin Hall's 1879 discovery \cite{Hall1879} that a current-carrying conductor develops a transverse voltage in the presence of a magnetic field. This arises as a simple consequence of the Lorentz force law \cite{Lorentz1895}; a charge $q$ moving in a static and spatially uniform magnetic field $\bm{B}$ experiences a force $q(\bm{v}/c) \times \bm{B}$, which deflects it in the direction perpendicular to its velocity $\bm{v}$ and the magnetic field, resulting in a transverse voltage across the sample. For a magnetic field of magnitude $B^z$ oriented along the $z$-direction, the Hall resistivity scales linearly with the field magnitude,
\begin{align}
    \rho_{H}^{xy} = R_{0} B^z,
    \nonumber
\end{align}
where the coefficient $R_0$ depends on material properties such as the carrier density \cite{Drude1900a, Drude1900b}. This Hall resistivity vanishes when the magnetic field is ``turned off," leaving only the usual longitudinal Ohmic flow of charge carriers that arises in any nonmagnetic conductor. 

But soon after the classical Hall effect was discovered, Edwin Hall observed a similar transverse voltage in the ferromagnetic phase of iron, even in the absence of any applied magnetic field \cite{Hall1880}. This \textit{anomalous Hall effect} was measured to be roughly ten times larger than that expected from the ordinary Hall effect. Experiments on several such magnetic conductors indicate that there is an additional term in the Hall resistivity proportional to the component $M^z$ of the magnetization that is parallel to the magnetic field \cite{Pugh1950,Pugh1953},
\begin{align}
    \rho_{H}^{xy} = R_{0} B^z + R_{s} M^z,
    \nonumber
\end{align}
where the ``anomalous coefficient" $R_s$ is found to depend in a subtle way on a number of material parameters, and also on the longitudinal resistivity. That there can develop a transverse ``Hall voltage" even in the absence of a magnetic field is not explained by the Lorentz force law, and the physical mechanism behind this effect has long been debated in the literature \cite{AHEcolloquium}. 

One of the first theoretical descriptions of the anomalous Hall effect was introduced by Karplus and Luttinger \cite{KarplusLuttinger1954}, who showed that electrons moving in a magnetic conductor under an external electric field can acquire an ``anomalous velocity" in a direction perpendicular to the electric field. This anomalous velocity results from the spin-orbit coupling between the orbital motion of itinerant electrons and the intrinsic magnetization of the crystal. However, many researchers soon pointed out that this ``intrinsic" process alone is insufficient to fully capture the experimental data \cite{Smit1955, Smit1958, Berger1970}, and two additional ``extrinsic" scattering mechanisms must also be considered. One such mechanism is ``skew-scattering," where the presence of both spin-orbit coupling and magnetic order leads to an asymmetric scattering of charge carriers off impurities, and thereby to an unbalanced transverse motion of these carriers \cite{Smit1955, Smit1958}. And there is another kind of asymmetric scattering called ``side-jump," in which these carriers experience an unbalanced electric field when approaching and leaving an impurity \cite{Berger1970}. Both of these \textit{extrinsic} scattering mechanisms generally exist in a ferromagnetic metal, and coexist with the intrinsic spin-orbit coupling mechanism of the Karplus-Luttinger theory \cite{Onoda2006, Sinitsyn2008}.

A more ``modern" picture emerged when this intrinsic mechanism was reformulated in geometric terms through the Berry curvature of the crystal's band structure \cite{Jungwirth2002,Yao2004,Xiao2010}. There, the part of the Hall conductivity tensor describing this intrinsic mechanism is rewritten using concepts from quantum geometry, and this ``intrinsic anomalous Hall effect" becomes one member of a broader family of geometric and topological response phenomena \cite{Berry1984,TKNN1982}. Haldane further emphasized that, in metals, the intrinsic part of the Hall conductivity, although written as an integral over occupied states, has an essential Fermi-surface character in metallic systems \cite{Haldane2004}; the points on the Fermi surface where the occupation is discontinuous prevent it from being topologically quantized as it is in the \textit{quantum} anomalous Hall effect \cite{Haldane1988,TKNN1982}. Around the same time, detailed analyses using Green functions and the Kubo formalism \cite{Kubo1957,Luttinger1958}, and others based on quantum kinetics \cite{Onoda2008}, were shown to provide mutually consistent descriptions of the anomalous Hall effect once the intrinsic and extrinsic mechanisms are treated in a common framework. 

Despite these advances, the anomalous Hall effect is most often studied directly at the level of transport functions, employing the conductivity tensor \cite{Kubo1957}, the quantum kinetic transport equations \cite{Onoda2008}, or diagrammatic calculations within linear response theory \cite{Luttinger1958}. However, in focusing only on the final form of the transport functions it is not clear how the charges and currents are behaving at the microscopic level when the electric field is applied. This is completely adequate for a DC calculation, since in this static limit charges rearrange themselves in such a way that the field in the interior of the material is screened \cite{Thomas1927,Fermi1927}. But when considering electric fields at finite frequency, there is an ongoing dynamical rearrangement of the charge-current distribution, and it is interesting to consider how the polarization of bound charges and the flow of itinerant electrons in this dynamical picture lead simultaneously to the optical polarization of bound charges \cite{Lorentz1895}, the metallic Drude response \cite{Drude1900a,Drude1900b}, and the anomalous Hall effect.

We revisit the anomalous Hall effect in magnetic conductors from this perspective. Our aim is not to propose a new mechanism for the anomalous Hall effect, but instead to investigate the dynamics of charges and currents at the microscopic level in terms of notions of polarization, magnetization, and free charges and currents. Beginning with a Hamiltonian field theory for the electronic degrees of freedom that is formulated within second-quantized quantum mechanics \cite{Kattan2026}, we define charge and current densities that can be decomposed into microscopic polarization and magnetization fields, and free charge and current densities, which can be associated respectively with ``bound" and ``free" charges in a physically intuitive way \cite{Mahon2019,Mahon2021}. We derive expressions for these fields in the ground state of the crystal, and then study how these expressions are modified when a spatially uniform, time-dependent electric field is applied. And by spatially averaging these microscopic fields, we calculate the corresponding macroscopic polarization and magnetization fields, and free charge and current densities.

The macroscopic current density that results involves a frequency-dependent conductivity tensor consisting of a sum of three terms, each of which can meaningfully be attributed to one of the polarization or free current responses. There is a ``Kubo term" that arises because of the polarization of bound charges induced by the electric field, analogous to the usual polarization response of a crystalline insulator \cite{Kattan2025}, except insofar as the band occupation in a metal changes across the Brillouin zone. The presence of a Fermi surface in metals also leads to a nonvanishing Drude term, describing the finite-frequency generalization of the usual Ohmic response of a metal to a DC electric field, and which is identified with the longitudinal part of the induced free current density. And the transverse part of this induced free current density is encoded precisely in the Hall term that describes the anomalous Hall effect. Very much in the spirit of the original Karplus-Luttinger theory \cite{AHEcolloquium}, we write the Kubo and Hall contributions to the conductivity tensor in terms of velocity matrix elements, which in our field theory involve a static vector potential and a corresponding magnetic field that encode the spontaneous magnetization of a magnetic conductor at the microscopic level.

We perform a numerical calculation of the Kubo, Drude, and Hall conductivities separately for the body-centered cubic phase of iron, the material in which the anomalous Hall effect was first discovered. From \textit{ab initio} and Wannier-interpolated band structures, we calculate each of these contributions individually. We find that at photon energies below 0.5 eV both the transverse and longitudinal responses are dominated by the Hall and Drude terms, while at higher frequencies the (interband) Kubo term becomes comparable in magnitude. The total frequency-dependent longitudinal response, comprising the Drude and Kubo conductivities, agrees qualitatively with previous experimental measurements \cite{PhysRevB.9.5056}. 

We begin in Sec.~\ref{Sec:Hamiltoniantheory} with a summary of the Hamiltonian field theory in terms of which the electron dynamics is studied. Because the Hamiltonian for the unperturbed system is cell-periodic, the electronic structure can be studied using elementary concepts from condensed matter physics such as Bloch's theorem and Wannier functions, and the electric field is included in the Hamiltonian through minimal coupling. The microscopic notions of polarization, magnetization, and free charges and currents are introduced in Sec.~\ref{Sec:Polarizationmagnetization}. Through a perturbative expansion of these quantities that is summarized in Appendix~\ref{Appendix:Structuredynamics}, we calculate the macroscopic current density that is induced by a spatially uniform electric field at finite frequency in Sec.~\ref{Sec:Dynamics}, leading to the long-wavelength conductivity tensor discussed above. Moreover, in Sec.~\ref{Sec:Samplecalculation} we provide a sample calculation of the optical conductivity for ferromagnetic iron, illustrating the interplay between the Hall, Drude, and Kubo terms in the conductivity tensor.

\section{Field theory}\label{Sec:Hamiltoniantheory}

We consider a bulk, nominally infinite crystal in $d$ dimensions, treated in the frozen-ion approximation where the ions in the crystal's lattice $\Gamma$ are fixed in space. We focus on ``magnetic conductors," which are metals that acquire a long-range magnetic order as they are cooled below the critical temperature of the corresponding phase transition. Working in the Heisenberg picture, the electronic degrees of freedom for spin-$1/2$ electrons are described by a pair of electron field operators $\hat{\psi}_{s}(\bm{x},t)$ indexed by $s \in \{\uparrow,\downarrow\}$, satisfying the equal-time anticommutation relations
\begin{align}
    \big[\hat{\psi}_{s}(\bm{x},t), \hat{\psi}_{s'}^{\dagger}(\bm{y},t)\big]_+ &= \delta_{ss'} \delta(\bm{x} - \bm{y}),\nonumber \\
    \big[\hat{\psi}_{s}(\bm{x},t), \hat{\psi}_{s'}(\bm{y},t)\big]_+ &= \big[\hat{\psi}_{s}^{\dagger}(\bm{x},t), \hat{\psi}_{s'}^{\dagger}(\bm{y},t)\big]_+ = 0,
    \label{anticommutationrelations}
\end{align}
ensuring that the electrons obey the correct Fermi-Dirac statistics associated with the Pauli exclusion principle. These are collected into a $2$-spinor field operator 
\begin{align}
    \hat{\psi}_0(\bm{x},t) \equiv \left(\mqty{\hat{\psi}_{\uparrow}(\bm{x},t) \\[2pt] \hat{\psi}_{\downarrow}(\bm{x},t)}\right).
\end{align}
Restricting ourselves to the independent-particle approximation, where Coulomb interactions between the electrons are treated only at the mean-field level, the crystal Hamiltonian is \cite{Duffspin,Kattan2026}
\begin{align}
    \hat{H}_0(t) = \int d\bm{x}\, \hat{\psi}_{0}^{\dagger}(\bm{x},t) \mathcal{H}_{0}(\bm{x}) \hat{\psi}_{0}(\bm{x},t),
    \label{H0}
\end{align}
involving the Hamiltonian density
\begin{align}
    \mathcal{H}_0(\bm{x}) = \frac{1}{2m}(\bm{\mathfrak{p}}(\bm{x}))^2 + \mathrm{V}_{\Gamma}(\bm{x}) - \frac{e\hbar}{2mc} \bm{\sigma}\cdot\bm{b}_{\mathrm{static}}(\bm{x})\\
    + \frac{\hbar^2}{4m^2 c^2} \bm{\sigma}\cdot\bm{\nabla}\mathrm{V}_{\Gamma}(\bm{x}) \times \bm{\mathfrak{p}}(\bm{x}),
    \label{H0density}
\end{align}
where we have introduced the vector $\bm{\sigma} = (\sigma^x, \sigma^y, \sigma^z)$ of Pauli matrices, along with the modified momentum operator
\begin{align}
    \bm{\mathfrak{p}}(\bm{x}) = \frac{\hbar}{i}\bm{\nabla} - \frac{e}{c} \bm{a}_{\mathrm{static}}(\bm{x}).
    \label{frakturp}
\end{align}
This is the first-quantized canonical momentum operator in the presence of a static vector potential $\bm{a}_{\mathrm{static}}(\bm{x})$ and corresponding magnetic field
\begin{align}
    \bm{b}_{\mathrm{static}}(\bm{x}) = \bm{\nabla} \times \bm{a}_{\mathrm{static}}(\bm{x}),
    \label{staticB}
\end{align}
both of which are periodic under translations in the crystal's lattice. These classical fields encode in an effective way the various microscopic exchange interactions that establish long-range magnetic order in the magnetic phase, and their presence describes the spontaneously broken time-reversal symmetry in the system's ground state. The electrostatic potential energy $\mathrm{V}_{\Gamma}(\bm{x})$ is also cell-periodic, and includes both the lattice potential due to the ions along with the usual Hartree correction within the independent-particle approximation. 

All of the quantities in Eq.~(\ref{H0density}) are cell-periodic, and therefore so too is the Hamiltonian density itself, implying that the solutions of the spectral equation
\begin{align}
    \mathcal{H}_0(\bm{x}) \psi_{n\bm{k}}(\bm{x}) = E_{n\bm{k}} \psi_{n\bm{k}}(\bm{x})
\end{align}
are the $2$-spinor Bloch energy eigenfunctions
\begin{align}
    \psi_{n\bm{k}}(\bm{x}) = \frac{1}{(2\pi)^{d/2}} e^{i\bm{k}\cdot\bm{x}} u_{n\bm{k}}(\bm{x}),
\end{align}
where the $2$-spinor Bloch functions
\begin{align}
    u_{n\bm{k}}(\bm{x}) = \left(\mqty{u_{n\bm{k}}^{\uparrow}(\bm{x}) \\[2pt] u_{n\bm{k}}^{\downarrow}(\bm{x})}\right)
\end{align}
are also cell-periodic. The Bloch wavevector $\bm{k}$ is a point in the $d$-dimensional Brillouin zone $\mathrm{BZ}^d$, and $n \in \mathbb{N}$ labels an energy band in the crystal's band structure. Because we are considering metals for which there exists a Fermi surface, which is the manifold of points $\bm{k}_F \in \mathrm{BZ}^d$ such that $E_{n\bm{k}_F} = E_F$ with $E_F$ being the Fermi energy, the crystal's band structure is gapless, having partially filled bands at the Fermi energy. Then the occupation factor 
\begin{align}
    f_{n\bm{k}} = \Theta(E_F - E_{n\bm{k}}),
    \label{Fermioccupation}
\end{align}
which is the zero-temperature limit of the Fermi-Dirac distribution, will depend explicitly on the Bloch wavevector $\bm{k}$ across the Brillouin zone. 

The perspective we introduce is based on a real-space formulation of electronic structure in crystals in terms of exponentially localized Wannier functions. To define these Wannier functions, we make a local unitary transformation of the Bloch functions,
\begin{align}
    u_{\alpha\bm{k}}(\bm{x}) = \sum_{n} U_{n\alpha}(\bm{k}) u_{n\bm{k}}(\bm{x}),
    \label{quasiBloch}
\end{align}
where the matrix $U_{n\alpha}(\bm{k})$ is chosen in such a way that the quasi-Bloch functions on the left-hand-side are globally smooth as functions of $\bm{k} \in \mathrm{BZ}^d$. Then the corresponding Wannier functions are defined by
\begin{align}
    W_{\alpha\bm{R}}(\bm{x}) = \sqrt{\Omega_{\mathrm{uc}}} \int_{\mathrm{BZ}^d} \frac{d\bm{k}}{(2\pi)^d}\, e^{i\bm{k}\cdot(\bm{x} - \bm{R})} u_{\alpha\bm{k}}(\bm{x}),
    \label{ELWF}
\end{align}
where $\Omega_{\mathrm{uc}}$ is the volume of the unit cell. In a topologically trivial \textit{insulator}, it is always possible to find a unitary transformation (\ref{quasiBloch}) of the occupied ``valence" bands in such a way that these Wannier functions will be exponentially localized. In a metal, however, where there is no meaningful separation of ``occupied valence bands" and ``unoccupied conduction bands" by a bandgap, this unitary transformation will generally involve the bands below and at the Fermi energy, as well as one or more unoccupied bands above $E_F$ that together constitute an isolated ``island" of bands in the crystal's band structure. This is also the case for Chern insulators, for example, as has been discussed in detail in other work \cite{albert2026linear}. 

We are interested in the response of magnetic conductors to time-dependent electric fields in the long-wavelength limit, and so we neglect any effects due to the associated time-dependent macroscopic magnetic fields. We also ignore local field corrections, taking the electric field to be the macroscopic electric field written in the usual way
\begin{align}
    \bm{E}(\bm{x},t) = - \bm{\nabla}\phi(\bm{x},t) - \frac{1}{c} \frac{\partial \bm{A}(\bm{x},t)}{\partial t},
    \label{Efield}
\end{align}
in terms of a scalar potential $\phi(\bm{x},t)$ and a vector potential $\bm{A}(\bm{x},t)$. We do not fix a gauge and instead keep the potentials general, except insofar as the vector potential is taken to be curl-free given that we are neglecting effects due to the magnetic field of the light. The electric field is accounted for in the Hamiltonian (\ref{H0}) by implementing the standard minimal-coupling prescription to the Hamiltonian density (\ref{H0density}). The electronic degrees of freedom are described by a new $2$-spinor electron field operator $\hat{\psi}(\bm{x},t)$ obeying the same anticommutation relations (\ref{anticommutationrelations}), and evolving in time under the Hamiltonian
\begin{align}
    \hat{H}(t) = \int d\bm{x}\, \hat{\psi}(\bm{x},t) \mathcal{H}_{\mathrm{mc}}(\bm{x},t) \hat{\psi}(\bm{x},t),
\end{align}
where the minimal-coupling Hamiltonian density is
\begin{align}
    \mathcal{H}_{\mathrm{mc}}(\bm{x},t) =&\; \frac{1}{2m}\big(\bm{\mathfrak{p}}_{\mathrm{mc}}(\bm{x},t)\big)^2 + \mathrm{V}_{\Gamma}(\bm{x}) + e\phi(\bm{x},t)\nonumber \\
    &+ \frac{\hbar}{4m^2 c^2} \bm{\sigma}\cdot\bm{\nabla}\mathrm{V}_{\Gamma}(\bm{x}) \times \bm{\mathfrak{p}}_{\mathrm{mc}}(\bm{x},t)\nonumber \\
    &- \frac{e\hbar}{2mc} \bm{\sigma}\cdot\bm{b}_{\mathrm{static}}(\bm{x}),
    \label{Hmcdensity}
\end{align}
involving the minimal-coupling momentum operator
\begin{align}
    \bm{\mathfrak{p}}_{\mathrm{mc}}(\bm{x},t) \equiv \bm{\mathfrak{p}}(\bm{x}) - \frac{e}{c} \bm{A}(\bm{x},t).
\end{align}
An analysis of local charge conservation using Noether's theorem indicates that the dynamics of electrons in a crystal can be described in terms of the electronic charge density operator
\begin{align}
    \hat{\rho}_e(\bm{x},t) = e \hat{\psi}^{\dagger}(\bm{x},t) \hat{\psi}(\bm{x},t),
    \label{chargeoperator}
\end{align}
and the current density operator
\begin{align}
    \hat{\bm{j}}_e(\bm{x},t) =&\; \frac{e}{2mc}\Big(\hat{\psi}^{\dagger}(\bm{x},t)\big(\bm{\mathfrak{p}}_{\mathrm{mc}}(\bm{x},t)\hat{\psi}(\bm{x},t)\big) + \mathrm{H.c.}\Big)\nonumber \\
    &+ \frac{e\hbar}{4m^2 c^2} \hat{\psi}^{\dagger}(\bm{x},t) \big(\bm{\sigma} \times \bm{\nabla}\mathrm{V}_{\Gamma}(\bm{x})\big)\hat{\psi}(\bm{x},t)\nonumber \\
    &+ \frac{e\hbar}{2m} \bm{\nabla} \times \big(\hat{\psi}^{\dagger}(\bm{x},t) \bm{\sigma}\hat{\psi}(\bm{x},t)\big),
    \label{currentoperator}
\end{align}
where the first line is the standard minimal-coupling current density, while the second and third lines are additional spin contributions due respectively to spin-orbit coupling and the spin magnetization in the crystal.

\section{Polarization and magnetization}\label{Sec:Polarizationmagnetization}

We study the physics of magnetic metals using a formalism based on microscopic notions of polarization and magnetization in extended systems. Beginning with the electronic charge density operator (\ref{chargeoperator}) and current density operator (\ref{currentoperator}), along with a static charge density $\rho_{\mathrm{ion}}(\bm{x})$ for the frozen ions in the crystal's lattice, we form the microscopic charge and current densities
\begin{align}
    \rho(\bm{x},t) &= \langle\hat{\rho}_e(\bm{x},t)\rangle + \rho_{\mathrm{ion}}(\bm{x}),\nonumber \\
    \bm{j}(\bm{x},t) &= \langle\hat{\bm{j}}_e(\bm{x},t)\rangle,
    \label{chargecurrentdensities}
\end{align}
involving expectation values of the electronic operators. Our strategy is to introduce \textit{microscopic} polarization and magnetization fields $(\bm{p},\bm{m})$, which together with microscopic free charge and current densities $(\rho_F,\bm{j}_F)$ satisfy
\begin{align}
    \rho(\bm{x},t) &= - \bm{\nabla} \cdot \bm{p}(\bm{x},t) + \rho_F(\bm{x},t),\nonumber \\
    \bm{j}(\bm{x},t) &= \frac{\partial \bm{p}(\bm{x},t)}{\partial t} + c \bm{\nabla} \times \bm{m}(\bm{x},t) + \bm{j}_F(\bm{x},t).
    \label{microscopicpolmag}
\end{align}
This kind of formalism is standard in the theory of optical response for atoms and molecules \cite{Healybook, Kattan2023}. Its extension to crystals was originally formulated for topologically trivial insulators \cite{Mahon2019}, and was later extended to nonmagnetic metals \cite{Mahon2021} and Chern insulators \cite{Mahon2023,Kattan2025}. We briefly summarize the central constructions below.

The starting point is a complete set of exponentially localized Wannier functions (\ref{ELWF}) for the unperturbed system. When an electric field (\ref{Efield}) is present, the treatment of the vector potential involves multiplying these Wannier functions by a generalized Peierls phase 
\begin{align}
    \Phi(\bm{x},\bm{R};t) = \frac{e}{\hbar c} \int d\bm{w}\, s^i(\bm{w};\bm{x},\bm{R}) A^i(\bm{w},t),
    \label{Peierlsphase}
\end{align}
involving a so-called ``relator" 
\begin{align}
    \bm{s}(\bm{w};\bm{x},\bm{R}) = \int_{C(\bm{x},\bm{R})} d\bm{z}\, \delta(\bm{w} - \bm{z}),
    \label{srelator}
\end{align}
which implements a line integral of the vector potential along a path $C(\bm{R},\bm{R}')$ connecting the lattice site $\bm{R}$ to the point $\bm{x}$ in space. The resulting ``modified" Wannier functions are 
\begin{align}
    W_{\alpha\bm{R}}'(\bm{x},t) = e^{i\Phi(\bm{x},\bm{R};t)} W_{\alpha\bm{R}}(\bm{x}),
    \label{modifiedWF}
\end{align}
which, in the long-wavelength approximation, will constitute a complete, orthonormal set of basis functions in terms of which the electron field operator can be expanded as
\begin{align}
    \hat{\psi}(\bm{x},t) = \sum_{\alpha\bm{R}} W_{\alpha\bm{R}}'(\bm{x},t) \hat{a}_{\alpha\bm{R}}(t),
    \label{fieldoperatorexpansion}
\end{align}
where the associated creation and annihilation operators satisfy the canonical anticommutation relations. Implementing this expansion in the charge density (\ref{chargeoperator}) and the current density (\ref{currentoperator}) operators, and taking the expectation values to form the microscopic charge and current densities (\ref{chargecurrentdensities}), those densities admit lattice decompositions of the form
\begin{align}
    \rho(\bm{x},t) &= \sum_{\bm{R}} \rho_{\bm{R}}(\bm{x},t),\nonumber \\
    \bm{j}(\bm{x},t) &= \sum_{\bm{R}} \bm{j}_{\bm{R}}(\bm{x},t),
    \label{sitechargecurrent}
\end{align}
where expressions for the ``site" charge and current densities can be found in Appendix~\ref{Appendix:Structuredynamics}.

One then introduces a polarization field $\bm{p}_{\bm{R}}(\bm{x},t)$ and a magnetization field $\bm{m}_{\bm{R}}(\bm{x},t)$ for each lattice site $\bm{R} \in \Gamma$ in the crystal, and the total microscopic polarization and magnetization fields are
\begin{align}
     \bm{p}(\bm{x},t) &= \sum_{\bm{R}} \bm{p}_{\bm{R}}(\bm{x},t),\nonumber \\
     \bm{m}(\bm{x},t) &= \sum_{\bm{R}} \bm{m}_{\bm{R}}(\bm{x},t).
\end{align}
Expressions for these can also be found in Appendix~\ref{Appendix:Structuredynamics}. And the site polarization and magnetization fields can be expanded in a series of generalized  electric and magnetic multipole moments, where, for example, the electric and magnetic dipole moments are 
\begin{align}
    \bm{\mu}_{\bm{R}}(t) &= \int d\bm{x}\, \bm{p}_{\bm{R}}(\bm{x},t),\nonumber \\
    \bm{\nu}_{\bm{R}}(t) &= \int d\bm{x}\, \bm{m}_{\bm{R}}(\bm{x},t).\label{dipolemoments}
\end{align}
But in a metallic system this is not the whole story, since there are mobile charge carriers that can flow between lattice sites when driven by an electric field. This redistribution of ``free" charges is described by the microscopic free charge and current densities in Eq.~(\ref{microscopicpolmag}). We model these free charges and currents through a kind of generalized lattice gauge theory, where at each lattice site $\bm{R} \in \Gamma$ we situate a site charge
\begin{align}
    Q_{\bm{R}}(t) \equiv \int d\bm{x}\, \rho_{\bm{R}}(\bm{x},t)
    \label{sitecharges}
\end{align}
and the free charge density is taken to be
\begin{align}
    \rho_F(\bm{x},t) = \sum_{\bm{R}} Q_{\bm{R}}(t) \delta(\bm{x} - \bm{R}).\label{microscopicfreecharge}
\end{align}
From the dynamical equation for the site charges,
\begin{align}
    \frac{d Q_{\bm{R}}(t)}{dt} = \sum_{\bm{R}'} I(\bm{R},\bm{R}';t),
\end{align}
we identify a collection of ``link currents" $I(\bm{R},\bm{R}';t)$ connecting the site charges between any pair of lattice sites, and we thereby define the free current density
\begin{align}
    \bm{j}_F(\bm{x},t) = \frac{1}{2} \sum_{\bm{R}\bm{R}'} \bm{s}(\bm{x};\bm{R},\bm{R}') I(\bm{R},\bm{R}';t),
    \label{microscopicfreecurrent}
\end{align}
involving the relator (\ref{srelator}). Notably, the free charge and current densities satisfy a continuity equation, so there remains a local conservation of charge. Analytical expressions for the site charges and link currents can be found elsewhere \cite{Mahon2019}.

Macroscopic notions of polarization and magnetization can be defined in terms of their microscopic counterparts above through an appropriate spatial-averaging procedure. Specifically, we choose an averaging function with a characteristic length scale $\Delta$ satisfying
\begin{align}
    a \ll \Delta \ll \lambda,
\end{align}
where $a$ is on the order of a lattice constant and $\lambda$ characterizes the typical range of variation of the electric field. The averaging function $w(\bm{x} - \bm{y})$ should be positive, spherically symmetric, and drop off continuously as the distance $\|\bm{x}-\bm{y}\| \to \infty$ with a decay length set by this characteristic length scale. Then the macroscopic polarization and magnetization fields are \cite{Mahon2020, Mahon2020a}
\begin{align}
    \bm{P}(\bm{x},t) &= \int d\bm{y}\, w(\bm{x}-\bm{y}) \bm{p}(\bm{y},t),\nonumber \\
    \bm{M}(\bm{x},t) &= \int d\bm{y}\, w(\bm{x}-\bm{y}) \bm{m}(\bm{y},t),\label{spatialaverage}
\end{align}
which together with the macroscopic free charge and current densities
\begin{align}
    \varrho_F(\bm{x},t) &= \int d\bm{y}\, w(\bm{x}-\bm{y}) \rho_F(\bm{y},t),\nonumber \\
    \bm{J}_F(\bm{x},t) &= \int d\bm{y}\, w(\bm{x}-\bm{y}) \bm{j}_F(\bm{y},t),\label{spatialaverage2}
\end{align}
satisfy the macroscopic identities
\begin{align}
    \varrho(\bm{x},t) &= - \bm{\nabla} \cdot \bm{P}(\bm{x},t) + \varrho_F(\bm{x},t),\nonumber \\
    \bm{J}(\bm{x},t) &= \frac{\partial \bm{P}(\bm{x},t)}{\partial t} + c \bm{\nabla} \times \bm{M}(\bm{x},t) + \bm{J}_F(\bm{x},t),
    \label{macroscopicfields}
\end{align}
where $\varrho(\bm{x},t)$ and $\bm{J}(\bm{x},t)$ are the macroscopic charge and current densities associated with the microscopic ones in Eq.~(\ref{chargecurrentdensities}) through the same averaging prescription.

\section{Dynamics}\label{Sec:Dynamics}

To study the anomalous Hall effect and its generalization to finite frequencies, we implement a perturbative expansion of the ``site" charge and current densities (\ref{sitechargecurrent}) in powers of the wavevector of the electromagnetic field. This leads to a similar expansion of the microscopic fields on the right-hand-side of Eq.~(\ref{microscopicpolmag}), and a perturbative description of the dynamics of the macroscopic polarization, magnetization, and free charges and currents in a crystal. Details concerning the calculation of these quantities can be found in Appendix~\ref{Appendix:Structuredynamics}.

We restrict ourselves to response in the linear regime, and emphasize that we make the \textit{long-wavelength approximation}, keeping only the terms that are independent of the wavevector of the macroscopic electric field. Since the magnetization contribution to the current density on the second line of Eq.~(\ref{macroscopicfields}) involves a spatial derivative, the only long-wavelength contributions to the current density come from the polarization field and the free current density. Then the macroscopic polarization and free current admit the perturbative expansions
\begin{align}
    \bm{P}(t) &= \bm{P}^{(0)} + \bm{P}^{(1)}(t) + \dots,\nonumber \\
    \bm{J}_F(t) &= \bm{J}_F^{(0)} + \bm{J}_F^{(1)}(t) + \dots,
\end{align}
and we keep only the zeroth-order part of the magnetization. These zeroth-order contributions, which are associated with the electronic structure of the unperturbed crystal in its zero-temperature ground state, are given in Appendix~\ref{Appendix:Structuredynamics}, and the higher-order contributions indicated by ellipses will be neglected. The first-order macroscopic charge density is easily shown to vanish, because the polarization field is uniform and the macroscopic free charge density vanishes, while the first-order macroscopic current density is
\begin{align}
    \bm{J}^{(1)}(t) = \frac{\partial \bm{P}^{(1)}(t)}{\partial t} + \bm{J}_F^{(1)}(t).
    \label{J1}
\end{align}
Introducing the standard Fourier-series decomposition
\begin{align}
    \bm{J}^{(1)}(\omega) = \sum_{\omega} \bm{J}^{(1)}(\omega) e^{-i\omega t},
\end{align}
and similarly for the first-order polarization field and free current density, this relation becomes
\begin{align}
    J^{(1)i}(\omega) = - i \omega P^{(1)i}(\omega) + J_F^{(1)i}(\omega).
    \label{longwavelengthresponse2}
\end{align}
The first-order contributions on the right-hand-side are given in Appendix~\ref{Appendix:Structuredynamics}. Upon combining these two contributions, the first-order current density induced by a uniform electric field takes the usual form
\begin{align}
    J^{(1)i}(\omega) = \sigma^{i\ell}(\omega) E^{\ell}(\omega),
    \label{longwavelengthresponse}
\end{align}
where the long-wavelength conductivity tensor is a sum of three contributions,
\begin{align}
    \sigma^{i\ell}(\omega) = \sigma_{\mathrm{K}}^{i\ell}(\omega) + \sigma_{\mathrm{D}}^{i\ell}(\omega) + \sigma_{\mathrm{H}}^{i\ell}.
    \label{conductivitytensor}
\end{align}

The first term, called the \textit{Kubo conductivity tensor}, describes the dominant response of the system at optical frequencies, and is given by
\begin{widetext}
\begin{align}
    \sigma_{\mathrm{K}}^{i\ell}(\omega) = - i\omega e^2 \hbar^2  \int_{\mathrm{BZ}^d} \frac{d\bm{k}}{(2\pi)^d} \sum_{mn} \frac{1}{(E_{m\bm{k}} - E_{n\bm{k}})^2} \frac{f_{nm,\bm{k}} v_{nm}^i(\bm{k}) v_{mn}^{\ell}(\bm{k})}{E_{m\bm{k}} - E_{n\bm{k}} - \hbar(\omega + i0^+)},
    \label{Kuboterm}
\end{align}
\end{widetext}
where the factor $f_{nm,\bm{k}} \equiv f_{n\bm{k}} - f_{m\bm{k}}$ is only nonzero for interband contributions. This tensor features the Cartesian components of the velocity matrix elements
\begin{align}
    \bm{v}_{nm}(\bm{k}) = \frac{1}{\Omega_{\mathrm{uc}}} \int_{\Omega} d\bm{x}\, u_{n\bm{k}}^{\dagger}(\bm{x}) \bm{v}(\bm{x}) u_{m\bm{k}}(\bm{x}),
    \label{velocitymatrixelements}
\end{align}
where the velocity operator
\begin{align}
    \bm{v}(\bm{x}) = \frac{1}{m}\bm{\mathfrak{p}}(\bm{x}) + \frac{\hbar}{4m^2 c^2} \bm{\sigma}\times \bm{\nabla}\mathrm{V}_{\Gamma}(\bm{x})
\end{align}
involves both the magnetic structure of the crystal, through the static vector potential in the momentum operator (\ref{frakturp}), and the spin-orbit coupling of the electrons. Notably, this Kubo conductivity tensor comes from the first-order polarization field in Eq.~(\ref{J1}), and has both longitudinal and transverse components, each of which contributes to both absorption and the optically induced polarization of bound charges in the crystal.  

The second term in the conductivity tensor (\ref{conductivitytensor}) describes the characteristic Drude response of a metal,
\begin{align}
    \sigma_{\mathrm{D}}^{i\ell}(\omega) = \frac{e^2}{\hbar} \frac{i}{\hbar(\omega + i0^+)} \int_{\mathrm{BZ}^d} \frac{d\bm{k}}{(2\pi)^d} \sum_n f_{n\bm{k}} \partial_{\ell} \partial_i E_{n\bm{k}},
    \label{Drudeterm}
\end{align}
This \textit{Drude conductivity tensor} is symmetric, arising from the longitudinal part of the free current response in Eq.~(\ref{J1}), and is the finite-frequency generalization of the usual Ohmic conductivity in the static limit for DC response. That this is a purely metallic contribution follows from the observation that it vanishes as a total derivative over the Brillouin zone when the occupation factor (\ref{Fermioccupation}) does not depend on the Bloch wavevector. That is, it vanishes for materials that are insulators. 

And the third term in the conductivity tensor (\ref{conductivitytensor}) is precisely the \textit{Hall conductivity}
\begin{align}
    \sigma_{\mathrm{H}}^{i\ell} = - e^2 \hbar \Im \int_{\mathrm{BZ}^d} \frac{d\bm{k}}{(2\pi)^d} \sum_{mn} \frac{ f_{nm,\bm{k}} v_{nm}^i(\bm{k}) v_{mn}^{\ell}(\bm{k})}{(E_{n\bm{k}} - E_{m\bm{k}})(E_{m\bm{k}} - E_{n\bm{k}})},
    \label{Hallterm}
\end{align}
which arises from the transverse part of the free current response in Eq. (\ref{J1}) and describes the anomalous Hall effect in ferromagnetic metals. Unlike the Kubo term (\ref{Kuboterm}), which is nonzero even in the limit of a time-reversal invariant and topologically trivial insulator, this Hall term vanishes when time-reversal symmetry holds. Indeed, if the static vector potential that features in the velocity matrix elements (\ref{velocitymatrixelements}) through the momentum operator (\ref{frakturp}) vanishes, then the integrand becomes odd under time reversal and thereby vanishes when integrated over the periodic Brillouin zone. From this microscopic perspective, the magnetic structure of the crystal, which is encoded in this static vector potential and the corresponding magnetic field (\ref{staticB}), is fundamental to the anomalous Hall response of a metal.

\section{Optical response of iron}\label{Sec:Samplecalculation}
\begin{figure}[b]
    \centering  \includegraphics[width=\linewidth]{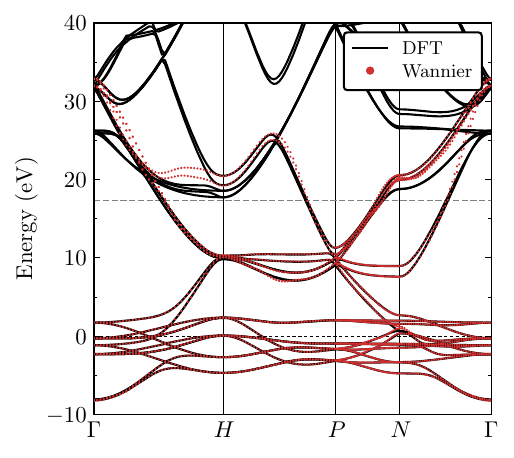}
    \caption{ The \textit{ab initio} and Wannier-interpolated band structure of bcc iron along high-symmetry path, in close agreement with X. Wang \textit{et al.} \cite{PhysRevB.74.195118}.  The Fermi energy is centered at 0 eV, and the horizontal dashed line at $17.37$ eV denotes the frozen energy window maximum used in the disentanglement step in the Wannier construction.}
\label{fig:structure}
\end{figure}

As an example, we 
evaluate the Kubo, Hall, and Drude contributions to the conductivity tensor in bulk bcc iron ($d = 3$).
First-principles density functional theory (DFT) calculations were performed using the Quantum ESPRESSO package \cite{Giannozzi_2009}. The fully relativistic Perdew-Burke-Ernzerhof (PBE) exchange-correlation functional was used \cite{PhysRevLett.77.3865},  together with norm-conserving pseudopotentials. Self-consistent calculations were carried out on a $\Gamma$-centered $\boldsymbol{k}$-grid of $16\times 16 \times
16$ with a convergence threshold of $10^{-8}$ eV. The non-self-consistent step in the DFT calculations was performed on a sparse $\boldsymbol{k}$-grid of $8\times 8 \times 8$. The in-plane lattice constant was fixed at $a=2.87\, \text{\AA}$ \cite{PhysRevB.74.195118}. As a post-processing step, 18 maximally localized Wannier functions were constructed using the Wannier90 code package \cite{MOSTOFI20142309,Pizzi_2020}. These Wannier functions were constructed from s, p, and d-orbitals near the Fermi level. The DFT band structure for bcc iron is shown in Fig.~\ref{fig:structure} alongside that obtained from Wannier interpolation. Excellent agreement is observed within the frozen energy window (grey dashed line around $17.37$ eV), and we closely match the results of X. Wang \textit{et al.} \cite{PhysRevB.74.195118}. Response calculations were implemented in the WannierBerri code package \cite{tsirkin_high_2021}, employing a dense $1000\times 1000\times 1000$ $\boldsymbol{k}$-grid and adaptive mesh refinement for Brillouin zone integration.
The bulk bcc Fe crystal structure belongs to the space group Im$\bar{3}$m (no. 229), where the symmetries enforce that the components of the total conductivity tensor in the plane orthogonal to the magnetization satisfy
\begin{align}
    \sigma^{xx}(\omega) &= \sigma^{yy}(\omega),\nonumber \\
    \sigma^{xy}(\omega) &= - \sigma^{yx}(\omega),
\end{align}
and, in the ferromagnetic phase $\sigma^{zz}(\omega) \neq\sigma^{xx}(\omega)$ due to the anisotropy of the ferromagnetic moments along the z-direction. Throughout this work, we have adopted the Gaussian unit system, in which the conductivity is expressed in inverse seconds. To convert these values to SI units, the conductivity must be multiplied by a factor of $4\pi \varepsilon_0= 1.11 \times 10^{-10} \, \text{s} \, (\Omega \, \text{m})^{-1}$ \cite{PhysRevB.82.035104}. The anomalous Hall conductivity of bcc Fe was calculated to be $756.17 \, (\Omega \, \text{cm})^{-1}$ using Eq.~(\ref{Hallterm}). We find that increasing the $\boldsymbol{k}$-grid size from $300\times 300 \times 300$ to $1000\times 1000\times 1000$, along with increasing the number of adaptive iterations to 40, changes the result by less than approximately $\pm 0.4\%$. This value is in close agreement ($<0.1\%$) with the value of $756.76 \, (\Omega \, \text{cm})^{-1}$ reported by X. Wang \textit{et al.} \cite{PhysRevB.74.195118}. For the Kubo conductivity tensor (\ref{Kuboterm}), we replace $i0^+$ with $i\eta$ where the broadening parameter $\eta=100$ meV, corresponding to the default smearing implemented in WannierBerri, which smooths the response while preserving relevant features. 
To calculate the finite-frequency Drude conductivity, we incorporate scattering effects by taking $0^+ \rightarrow \tau^{-1}$ in Eq.~(\ref{Drudeterm}) \cite{10.1093/oso/9780198566335.003.0016}, where $\tau$ denotes the scattering time. The frequency-dependent Drude conductivity can then be written as 
\begin{equation}
    \sigma_{\mathrm{D}}^{i\ell}(\omega)=\frac{\sigma^{i \ell}_{\text{Ohm}}}{(1-i\omega \tau)},
\end{equation}
where $\sigma^{i \ell}_{\text{Ohm}}$ is the static limit ($\omega =0$) Drude response. We take $\tau = $ 8 fs, consistent with the estimated mean relaxation time extracted from the DC conductivity of 100 nm iron films \cite{10.1063/1.5142479,PhysRevB.82.035104}. This yields $\sigma^{xx}_{\text{Ohm}} =7.062 \times 10^6 \, (\Omega \, \text{m})^{-1}$, in close agreement with the experimentally reported DC conductivity of approximately $8 \times 10^6 \, (\Omega \, \text{m})^{-1}$ in 100 nm films \cite{10.1063/1.5142479}. 
\begin{figure*}[t]
    \centering  \includegraphics[width=\textwidth]{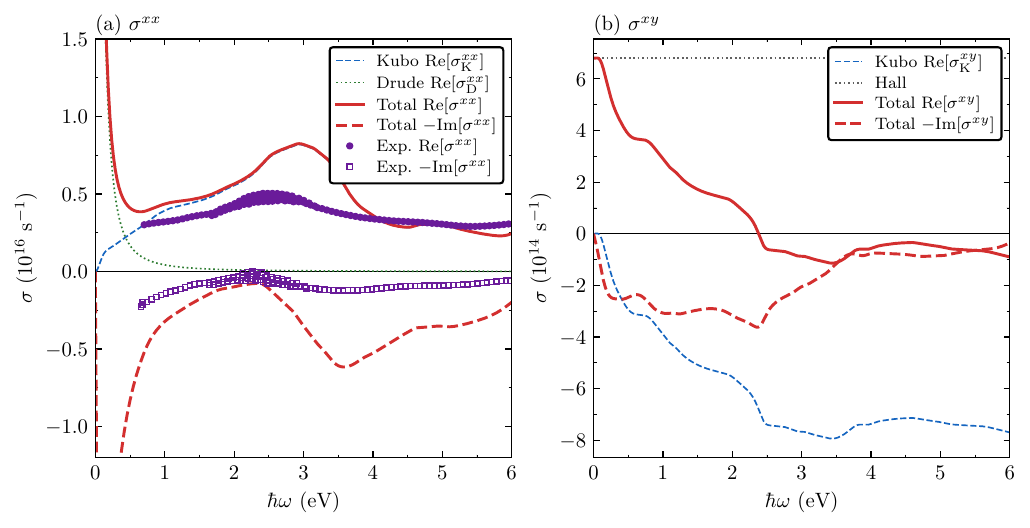}
    \caption{The frequency-dependent total conductivity of bcc iron along the (a) longitudinal and (b) transverse directions, which includes the sum of the Hall, Drude, and Kubo conductivity tensors. The experimental measurements of the optical conductivity of iron from Johnson and Christy are shown in (a) in purple at room temperature \cite{PhysRevB.9.5056}. }
\label{fig:totalConductivity}
\end{figure*}
 In Fig.~\ref{fig:totalConductivity}, we present the total longitudinal and transverse conductivities, including all three contributions, and find good qualitative agreement with the room temperature measurements found in Johnson and Christy \cite{PhysRevB.9.5056}. The crystal symmetries of bcc Fe restrict the Drude contribution to the diagonal components of the conductivity tensor, where it is the leading contribution at low frequencies. More precisely, when the energy exceeds $0.5$ eV, the Kubo (interband) contributions become the dominant term in the longitudinal response. A similar crossover occurs in the transverse response, where the anomalous Hall conductivity dominates at low energies below $2.5$ eV before interband contributions take over. The peak near 3 eV in the real part of the longitudinal conductivity is consistent in position and magnitude with previous zero temperature DFT calculations \cite{gpk7-gr7q,PhysRevB.82.035104}. In contrast, the experimental peak appears at approximately 2.5 eV. This discrepancy is attributed to finite-temperature effects in the room-temperature experiment performed by Johnson and Christy, which have been shown to redshift spectral features by reducing the exchange splitting caused by thermal demagnetization \cite{gpk7-gr7q}. In previous \textit{ab initio} calculations, the conductivity was calculated in the absence of the Drude contribution; in that case, we find that the imaginary part of the longitudinal conductivity matches the result of Silber \textit{et al.} \cite{PhysRevB.100.064403}.

\section{Conclusion}\label{Sec:Conclusion}

We have revisited the anomalous Hall effect in metals, and its generalization to finite frequencies, from the perspective of microscopic polarization and magnetization fields, and free charge and current densities. Starting with a second-quantized Hamiltonian for spin-$1/2$ electrons in a magnetically ordered crystal, where this magnetic order is also described at the microscopic level in terms of a static and cell-periodic vector potential (and an associated magnetic field), we calculated the electronic charge and current density operators in the presence of a time-dependent and spatially uniform electric field. The expectation values of these operators, along with a static charge density for the ions, defined microscopic charge and current densities that could be decomposed into a sum of contributions from each lattice site in the crystal. These ``site" charge and current densities are used to define site polarization and magnetization fields, which, when summed over the lattice, constitute the microscopic polarization and magnetization fields that, together with an appropriate pair of free charge and current densities, constitute the fundamental quantities of our approach. 

Upon spatially averaging these microscopic fields, and restricting ourselves to linear response in the long-wavelength limit, we found that the macroscopic current density induced by the electric field involves a conductivity tensor that naturally separates into a sum of three distinct terms. There is a Kubo term arising from the polarization response to the electric field, which describes the optical polarization of ``bound charges" in the medium. The metallic transport is described by the Drude conductivity tensor that is associated with the Fermi-surface geometry, and is the finite-frequency generalization of the usual Ohmic current in the static response of a metal. This Drude term comes from the longitudinal part of the free current response, while the transverse response was encoded in the static Hall conductivity that describes precisely the anomalous Hall effect, clarifying in a direct way how the Hall current is embedded in the transverse dynamics of mobile charge carriers. Much in the spirit of the original Karplus-Luttinger theory for the static anomalous Hall response, the Hall conductivity and the finite-frequency Kubo conductivity are expressed in terms of velocity matrix elements, which in our field theory also contain the static vector potential and a spin-orbit coupling piece, making explicit the role played by the broken time-reversal symmetry of the magnetic ground state.

Returning to the original material in which the anomalous Hall effect was first observed, we performed numerical calculations of the long-wavelength response of the ferromagnetic body-centered cubic phase of iron, evaluating the Kubo, Drude, and Hall contributions separately using \textit{ab initio} and Wannier-interpolated electronic-structure calculations. We find the Hall and Drude responses to be $756.17 \, (\Omega \, \text{cm})^{-1}$ and $70,621.84 \, (\Omega \, \text{cm})^{-1}$, respectively, in good agreement with previous experimental and numerical studies \cite{PhysRevB.74.195118,10.1063/1.5142479}. Our total conductivity, including the Drude contributions, exhibits the same qualitative features as previous experiments on bulk bcc iron \cite{PhysRevB.9.5056}.

The approach we have taken here does not introduce a new mechanism for the anomalous Hall effect; rather, it casts the known response of a magnetic conductor into what we believe to be a physically useful picture based on notions of polarization, magnetization, and free charges and currents in crystalline media. In doing so, we are able to study how each part of the response arises at the microscopic level. In this sense, our formalism provides a complementary description to the more standard quantum geometric and transport-based analyses by demonstrating how the optical polarization of bound charges, the metallic Drude response, and the anomalous Hall effect arise within a common description of the charge-current dynamics. The perturbative response calculations done here in the long-wavelength approximation can be straightforwardly extended to allow for spatially inhomogeneous electric and magnetic fields, and even to the nonlinear optical regime.



\begin{acknowledgments}
    This work was supported by the Natural Sciences and Engineering Research Council of Canada (NSERC). Computations were performed on the Trillium cluster hosted by the Digital Research Alliance of Canada. MA is supported by the Ontario Graduate Scholarship (OGS).
\end{acknowledgments}


\appendix


\section{Electronic structure and dynamics}\label{Appendix:Structuredynamics}

In this Appendix we provide some further details concerning the polarization and magnetization fields that we have used to study the electronic structure and response of a magnetic metal. Then we give expressions for the ground-state polarization and magnetization in these systems, and also for the first-order polarization and free current density that appear in Eq.~(\ref{J1}), which combine to yield the long-wavelength conductivity tensor (\ref{conductivitytensor}).

\subsection{Site quantities}

Beginning with the field operator expansion (\ref{fieldoperatorexpansion}) in terms of the modified Wannier functions (\ref{modifiedWF}) and their associated creation and annihilation operators, we substitute this expansion into the definitions (\ref{chargecurrentdensities}) of the microscopic charge and current densities. This leads to the lattice decompositions (\ref{sitechargecurrent}) into ``site" charge and current densities taking the generic form
\begin{align}
    \rho_{\bm{R}}(\bm{x},t) = \sum_{\alpha\beta\bm{R}'\bm{R}''} \rho_{\beta\bm{R}';\alpha\bm{R}''}(\bm{x},\bm{R};t) \eta_{\alpha\bm{R}'';\beta\bm{R}'}(t),\nonumber \\
    \bm{j}_{\bm{R}}(\bm{x},t) = \sum_{\alpha\beta\bm{R}'\bm{R}''} \bm{j}_{\beta\bm{R}';\alpha\bm{R}''}(\bm{x},\bm{R};t) \eta_{\alpha\bm{R}'';\beta\bm{R}'}(t),
    \label{siterhoj}
\end{align}
involving ``generalized site-quantity matrix elements" of the charge and current densities in the basis of modified Wannier functions, together with the \textit{single-particle density matrix} $\eta_{\alpha\bm{R}'';\beta\bm{R}'}(t)$, both of which are discussed in detail elsewhere \cite{Mahon2019, Mahon2021}. 

In terms of the ``site" charge density, the site polarization field at a lattice site $\bm{R} \in \Gamma$ is taken to be
\begin{align}
    p_{\bm{R}}^i(\bm{x},t) = \int d\bm{y}\, s^i(\bm{x};\bm{y},\bm{R}) \rho_{\bm{R}}(\bm{y},t),\label{micropolarization}
\end{align}
involving the relator (\ref{srelator}), while the site magnetization field is a sum of two terms, 
\begin{align}
    \bm{m}_{\bm{R}}(\bm{x},t) = \bar{\bm{m}}_{\bm{R}}(\bm{x},t) + \tilde{\bm{m}}_{\bm{R}}(\bm{x},t),
\end{align}
the first of which involves the ``site" current density (\ref{siterhoj}) and is taken to be
\begin{align}
    \bar{m}_{\bm{R}}^i(\bm{x},t) = \frac{1}{c} \int d\bm{y}\, \alpha^{ij}(\bm{x};\bm{y},\bm{R}) j_{\bm{R}}^j(\bm{y},t),
\end{align}
where we have introduced another ``relator" \cite{Mahon2019}
\begin{align}
    \alpha^{ij}(\bm{x};\bm{y},\bm{R}) = \varepsilon^{imn} \int_{C(\bm{y},\bm{R})} dz^m\, \frac{\partial z^n}{\partial y^j} \delta(\bm{x}-\bm{z}).\label{relatoralpha}
\end{align}
The second term is an ``itinerant" contribution, arising because there is a local non-conservation of charge at each lattice site separately, and is given by
\begin{align}
    \tilde{m}_{\bm{R}}^i(\bm{x},t) = \frac{1}{c} \int d\bm{y}\, \alpha^{ij}(\bm{x};\bm{y},\bm{R}) \tilde{j}_{\bm{R}}^j(\bm{y},t),
\end{align}
involving an itinerant ``site" current density $\tilde{\bm{j}}_{\bm{R}}(\bm{x},t)$ \cite{Mahon2019}. Similar expressions can be derived for the site charges in the free charge density (\ref{microscopicfreecharge}) and the link currents in the free current density (\ref{microscopicfreecurrent}). 

Choosing a straight-line path for the relators in Eqs.~(\ref{srelator}) and (\ref{relatoralpha}), and implementing a formal Taylor expansion of the resulting expressions, one obtains generalized electric and magnetic multipole expansions
\begin{align}
    p_{\bm{R}}^i(\bm{x},t) &= \mu_{\bm{R}}^i(t) \delta(\bm{x}-\bm{R}) - q_{\bm{R}}^{ij}(t) \frac{\partial \delta(\bm{x}-\bm{R})}{\partial x^j} + \dots,\nonumber \\
    m_{\bm{R}}^i(\bm{x},t) &= \nu_{\bm{R}}^i(t) \delta(\bm{x}-\bm{R}) + \dots,
\end{align}
where the site electric and magnetic dipole moments are given in Eq.~(\ref{dipolemoments}), and the second term on the first line involves the electric quadrupole moment, which we neglect here. Following the spatial averaging procedure we discuss in Sec.~\ref{Sec:Polarizationmagnetization}, the corresponding macroscopic fields can then be calculated.
\\

\subsection{Ground state}

Suppose there is no electric field and the metal is in its electronic ground state at zero temperature, described by the Fermi occupation factor (\ref{Fermioccupation}). Following the steps discussed above for this ``unperturbed system,"  we calculate the macroscopic polarization and magnetization, which are static and spatially uniform in the ground state. 

Focusing on bulk metals in $d = 3$ dimensions, the macroscopic polarization is found to be
\begin{align}
    P^{(0)i} = e \int_{\mathrm{BZ}^3} \frac{d\bm{k}}{(2\pi)^3} \sum_n f_{n\bm{k}} \Big(\xi_{nn}^i(\bm{k}) + \mathcal{W}_{nn}^i(\bm{k})\Big) + P_{\mathrm{ion}}^i,
    \label{unperturbedP}
\end{align}
where the first term is the electronic contribution, involving the diagonal components of the non-Abelian Berry connection
\begin{align}
	\xi_{mn}^i(\bm{k}) = \frac{i}{\Omega_{\text{uc}}} \int_{\Omega} d\bm{x}\, u_{m\bm{k}}^{\dagger}(\bm{x})\frac{\partial u_{n\bm{k}}(\bm{x})}{\partial k^i},
	\label{Berryconnection}
\end{align}
and also the Hermitian matrix \cite{VanderbiltBook}
\begin{align}
	\mathcal{W}^i_{mn}(\bm{k})\equiv i\sum\limits_{\alpha}\big(\partial_i U_{m\alpha}(\bm{k})\big)U^\dagger_{\alpha n}(\bm{k}),
    \label{W}
\end{align}
which encodes a kind of ``gauge freedom" in the choice of Wannier functions (\ref{ELWF}) through the unitary matrices in Eq.~(\ref{quasiBloch}). And the second term in Eq.~(\ref{unperturbedP}) is the contribution due to the frozen ions. This is in agreement with previous work for spinless electrons, and constitutes an extension to metals of the macroscopic polarization found in the ``modern theory of polarization" for insulating systems, as is evident from the $\bm{k}$-dependent occupation factor (\ref{Fermioccupation}).

The macroscopic magnetization is found to be a sum of three terms,
\begin{align}
    M^{(0)i} = M_{\mathrm{metal}}^{(0)i} + M_{\mathrm{magnetic}}^{(0)i} + M_{\mathrm{spin}}^{(0)i},
\end{align}
where the first term is independent of the magnetic order and persists for nonmagnetic metals; it is given by
\begin{widetext}
\begin{align}
    M_{\mathrm{metal}}^{(0)i} = \frac{e}{2\hbar c} \varepsilon^{iab}  \int_{\text{BZ}^3}\frac{d\bm{k}}{(2\pi)^3} \sum_{n} f_{n\bm{k}} E_{n\bm{k}} \partial_a\xi^b_{nn}(\bm{k}) - \frac{e}{2\hbar c} \varepsilon^{iab}  \int_{\text{BZ}^3}\frac{d\bm{k}}{(2\pi)^3} \sum_{mn} f_{n\bm{k}} E_{m\bm{k}} \Im\big(\xi^a_{nm}(\bm{k})\xi^b_{mn}(\bm{k})\big),
\end{align}
which is the metallic generalization of the ``atomic-like" magnetization that is found in the ``modern theory of magnetization" for insulators. The second term arises due to the magnetic structure of the system
\begin{align}
    M_{\mathrm{magnetic}}^{(0)i} = \frac{e}{2\hbar c} \varepsilon^{iab} \int_{\text{BZ}^3}\frac{d\bm{k}}{(2\pi)^3}   \sum_{mn} f_{nm,\bm{k}} E_{n\bm{k}}\Im\big(\mathcal{W}^a_{nm}(\bm{k})\mathcal{W}^b_{mn}(\bm{k})\big),
\end{align}
\end{widetext}
and vanishes for time-reversal symmetric media. It is sensitive to the magnetic order, and is also found in magnetic Chern insulators \cite{Mahon2023}, for example, except there the occupation factor (\ref{Fermioccupation}) does not depend on $\bm{k} \in \mathrm{BZ}^d$. Importantly, this term depends on the choice of unitary matrix in Eq.~(\ref{quasiBloch}), and is therefore not directly physically measurable. And the third term in the macroscopic magnetization is an explicit spin contribution, given by
\begin{align}
    M_{\mathrm{spin}}^{(0)i} = \frac{e}{mc} \int_{\mathrm{BZ}^d} \frac{d\bm{k}}{(2\pi)^d} \sum_n f_{n\bm{k}} S_{nn}^i(\bm{k}),
\end{align}
involving the diagonal components of the spin matrix elements
\begin{align}
    S_{mn}^i(\bm{k}) = \frac{\hbar}{2} \frac{1}{\Omega_{\mathrm{uc}}} \int_{\Omega} d\bm{x}\, u_{m\bm{k}}^{\dagger}(\bm{x}) \sigma^i u_{n\bm{k}}(\bm{x}).
\end{align}
The macroscopic free charge and current densities are found to vanish in the unperturbed ground state, as expected, as do the total charge and current densities on the left-hand-side of Eq.~(\ref{macroscopicfields}).

\subsection{Linear response}

When an electric field is present, there are modifications to both the site-quantity matrix elements and the single-particle density matrix that together form the ``site" charge and current densities (\ref{sitechargecurrent}). These modifications are treated perturbatively, where, for example, the single-particle density matrix admits an expansion of the form \cite{Mahon2020, Mahon2020a}
\begin{equation}
    \eta_{\alpha\bm{R}'';\beta\bm{R}'}(t) = \eta_{\alpha\bm{R}'';\beta\bm{R}'}^{(0)} + \eta_{\alpha\bm{R}'';\beta\bm{R}'}^{(1)}(t) + \dots.\label{expansionSPDM}
\end{equation}
The superscript ``$(0)$" indicates the contribution that is independent of the electric field, corresponding to the ground-state quantities discussed above, while the superscript ``$(1)$" denotes the contribution that is linear in the electric field, and so on. There are similar expansions of the generalized site-quantity matrix elements in Eq.~(\ref{sitechargecurrent}), since these matrix elements involve the vector potential for the electric field through the generalized Peierls phase (\ref{Peierlsphase}). This leads to an expansion of the ``site" charge and current densities themselves, and subsequently an expansion of the site polarization and magnetization fields. 

We work in the long-wavelength approximation, where any dependence of the response tensors on the wavevector of the electric field is neglected. Because the magnetization part of the macroscopic current density in Eq.~(\ref{macroscopicfields}) is always accompanied by a spatial derivative, it does not contribute to the long-wavelength response, while for time-dependent electric fields there are contributions coming from the first-order polarization field and free current density. Writing the first-order contribution to the macroscopic polarization field in the form
\begin{align}
    P^{(1)i}(\omega) = \chi_P^{i\ell}(\omega) E^{\ell}(\omega),
\end{align}
the susceptibility tensor $\chi_{P}^{i\ell}(\omega)$ is found to be
\begin{align}
    \chi_P^{i\ell}(\omega) =&\; e^2 \int_{\mathrm{BZ}^d} \frac{d\bm{k}}{(2\pi)^d} \sum_{mn}  \frac{f_{nm,\bm{k}}\xi_{nm}^i(\bm{k})\xi_{mn}^{\ell}(\bm{k})}{E_{m\bm{k}} - E_{n\bm{k}} - \hbar(\omega + i0^+)}\nonumber \\
    &+ e^2 \int_{\mathrm{BZ}^d} \frac{d\bm{k}}{(2\pi)^d} \sum_{mn} \frac{f_{nm,\bm{k}}\mathcal{W}_{nm}^i(\bm{k})\xi_{mn}^{\ell}(\bm{k})}{E_{m\bm{k}} - E_{n\bm{k}} - \hbar(\omega + i0^+)}\nonumber \\
    &+ i e^2 \int_{\mathrm{BZ}^d} \frac{d\bm{k}}{(2\pi)^d} \sum_n \frac{f_{n\bm{k}}\partial_{\ell}\big(\xi_{nn}^i(\bm{k}) + \mathcal{W}_{nn}^{i}(\bm{k})\big)}{\hbar(\omega + i0^+)}.
    \label{chiP}
\end{align}
This first-order polarization is spatially uniform, because the electric field is spatially uniform, and therefore it does not contribute to the total charge density on the first line of Eq.~(\ref{macroscopicfields}). Also, the macroscopic free charge density, which in this long-wavelength approximation takes the generic form 
\begin{align}
    \varrho_F(t) = \frac{e}{\Omega_{\text{uc}}} \sum_{\alpha} \eta_{\alpha\bm{R};\alpha\bm{R}}(t)
\end{align}
for any fixed $\bm{R} \in \Gamma$, can be shown to vanish, and hence there is no induced macroscopic charge density in the medium. But there will be an induced free current density, which in this long-wavelength approximation is
\begin{align}
    J_F^i(t) = \frac{1}{2\Omega_{\text{uc}}} \sum_{\bm{R}'} \big(R^i - R'^i\big) I(\bm{R},\bm{R}';t),
\end{align}
for any fixed $\bm{R} \in \Gamma$. Following a perturbative expansion, analogous to Eq.~(\ref{expansionSPDM}), of the link currents
\begin{align}
    I(\bm{R},\bm{R}';t) = I^{(0)}(\bm{R},\bm{R}') + I^{(1)}(\bm{R},\bm{R}';t) + \dots,
    \label{linkexpansion}
\end{align}
one finds the zeroth-order term vanishes, consistent with the vanishing of the free current density in the unperturbed ground state. But there is a nonzero first-order part of the link current, which leads to an induced free current density that can be written as
\begin{align}
    J_F^{(1)i}(\omega) = \sigma_F^{i\ell}(\omega) E^{\ell}(\omega),
\end{align}
where the free-current conductivity tensor is found to be
\begin{align}
    \sigma_F^{i\ell}(\omega) =&\; i \omega e^2 \int_{\mathrm{BZ}^d} \frac{d\bm{k}}{(2\pi)^d} \sum_{mn} \frac{f_{nm,\bm{k}} \mathcal{W}_{nm}^i(\bm{k}) \xi_{mn}^{\ell}(\bm{k})}{E_{m\bm{k}} - E_{n\bm{k}} - \hbar (\omega + i0^+)}\nonumber \\
    &+ \frac{e^2}{\hbar} \frac{i}{\hbar(\omega + i0^+)} \int_{\mathrm{BZ}^d} \frac{d\bm{k}}{(2\pi)^d} \sum_n f_{n\bm{k}} \partial_{\ell} \partial_i E_{n\bm{k}}\nonumber \\
    &- \frac{e^2}{\hbar} \int_{\mathrm{BZ}^d} \frac{d\bm{k}}{(2\pi)^d} \sum_{n} f_{n\bm{k}} \partial_i\big(\xi_{nn}^{\ell}(\bm{k}) + \mathcal{W}_{nn}^{\ell}(\bm{k})\big)\nonumber \\
    &- \frac{e^2}{\hbar} \int_{\mathrm{BZ}^d} \frac{d\bm{k}}{(2\pi)^d} \sum_{mn} f_{nm,\bm{k}} \Im\big(\mathcal{W}_{nm}^{\ell}(\bm{k})\mathcal{W}_{mn}^i(\bm{k})\big).
    \label{sigmaF}
\end{align}
Using the linear relation (\ref{J1}), the long-wavelength conductivity tensor is obtained from
\begin{align}
    \sigma^{i\ell}(\omega) = - i\omega\chi_P^{i\ell}(\omega) + \sigma_F^{i\ell}(\omega),
\end{align}
where, after some algebraic simplifications, we find
\begin{align}
    \sigma^{i\ell}(\omega) = &- i\omega e^2 \int_{\mathrm{BZ}^d} \frac{d\bm{k}}{(2\pi)^d} \sum_{mn} \frac{f_{nm,\bm{k}}\xi_{nm}^i(\bm{k}) \xi_{mn}^{\ell}(\bm{k})}{E_{m\bm{k}} - E_{n\bm{k}} - \hbar(\omega + i0^+)}\nonumber \\
    &+ \frac{e^2}{\hbar} \frac{i}{\hbar(\omega + i0^+)} \int_{\mathrm{BZ}^d} \frac{d\bm{k}}{(2\pi)^d} \sum_n f_{n\bm{k}} \partial_{\ell} \partial_i E_{n\bm{k}}\nonumber \\
    &- \frac{ie^2}{\hbar} \int_{\mathrm{BZ}^d} \frac{d\bm{k}}{(2\pi)^d}  \sum_{mn} f_{nm,\bm{k}} \xi_{nm}^i(\bm{k}) \xi_{mn}^{\ell}(\bm{k}),
\end{align}
an expression that is independent of how the Bloch functions are chosen. The first line, which arises from the polarization response (\ref{chiP}), is equivalent to the Kubo conductivity tensor (\ref{Kuboterm}), while the second and third terms, arising from the longitudinal and transverse parts of the free current response (\ref{sigmaF}), are the Drude (\ref{Drudeterm}) and Hall (\ref{Hallterm}) conductivities, respectively. Notably, when time-reversal symmetry is restored the Hall term is easily shown to vanish, leaving only the Kubo and Drude conductivities, as expected for the optical response of a nonmagnetic metal.

\bibliography{Bibliography.bib}

\end{document}